# Ceramic Methyltrioxorhenium

Rudolf Herrmann,[a] Klaus Tröster,[a] Georg Eickerling,[a] Christian Helbig,[a] Christoph Hauf,[a] Robert Miller,[a] Franz Mayr,[a] Hans–Albrecht Krug von Nidda,[b] Ernst-Wilhelm Scheidt,[a] Wolfgang Scherer[a]*

[a] *Chemische Physik und Materialwissenschaften, Universität Augsburg, Universitätsstr. 1, D-86159 Augsburg, Germany*
[b] *Experimentalphysik V, Elektronische Korrelationen und Magnetismus, Universität Augsburg, Universitätsstr. 1, D-86159 Augsburg*

**Abstract**

The metal oxide polymeric methyltrioxorhenium $[(CH_3)_xReO_3]_\infty$ is an unique representative of a layered inherent conducting organometallic polymer which adopts the structural motifs of classical perovskites in two dimensions (2D) in form of methyl-deficient, corner-sharing $ReO_5(CH_3)$ octahedra. In order to improve the characteristics of polymeric methyltrioxorhenium with respect to its physical properties and potential usage as an inherent-conducting polymer we tried to optimise the synthetic routes of polymeric modifications of **1** to obtain a sintered ceramic material, denoted ceramic MTO. *Ceramic* MTO formed in a solvent-free synthesis via auto-polymerisation and subsequent sintering processing displays clearly different mechanical and physical properties from *polymeric* MTO synthesised in aqueous solution. Ceramic MTO is shown to display activated Re-C and Re=O bonds relative to MTO. These electronic and structural characteristics of ceramic MTO are also reflected by a different chemical reactivity compared with its monomeric parent compound. First examples of the unprecedented reactivity of ceramic MTO in the field of amine oxidations are shown – results which warrant further exploitation.

## 1. Introduction

Methyltrioxorhenium (MTO) **1** is the prototype of an organometallic oxide with the metal in its highest oxidation state. Many derivatives, mainly base adducts[1] which form due to its strong Lewis-acidic properties, and applications in catalysis (e.g. in the field of olefin metathesis and oxidation reactions)[2] were reported. Besides the manifold usage of **1** in homogeneous oxidation catalysis recent studies also highlight its capability to act as precursor in materials synthesis of nanostructured $ReO_2$[3] or mixed valent $MM'O_3$ oxides (M, M' = transition metal),[4] respectively. For example, mixed $W_{1-x}Re_xO_3$ oxides which were hitherto only accessible via high temperature and high pressure routes (65 kbar and 1200°C)[5] could be obtained by an oxolation process employing $Na_6[H_2W_{12}O_{40}]$ and **1** at mild condition (100°C) in dilute HCl.[4] When the potential of **1** as a precursor in CVD of rhenium oxides was evaluated, it was observed during DSC experiments that heating of MTO to 120-130°C, i.e. above its melting point (106°C), does not lead to a simple decomposition, but to the formation of a copper-coloured solid within 1-2 hours. The resulting material was called "water-free *poly*-MTO",[6] in analogy to conventional "*poly*-MTO" **2**, a bronze-coloured water-containing conducting material which forms when MTO is heated in aqueous solutions to 80°C.[7] A striking difference between the two materials is the hardness of water-free *poly*-MTO which forms brittle pieces while water-intercalated **2** is a soft, graphite-like material. However, both *poly*-MTO modifications share several common properties; e.g. solubility in aqueous $H_2O_2$ which leads to peroxo complexes of **1**,[7b] and an intrinsic thermodynamic instability which manifests in a pressure-induced depolymerisation. Hence, when KBr pellets are pressed for IR measurements, the material responds to the mechanical stress with slow formation of $ReO_3$ and single crystals of monomeric MTO.[7] Especially, the pressure sensitivity of **2** is remarkable: even subtle pressure gradients during standard filtering and washing procedures may be sufficient to initiate this depolymerisation reaction of raw *poly*-MTO. Such pressure-induced depolymerisation reaction in combination with the pronounced hygroscopic character of **2** and water-free *poly*-MTO limits its potential usage as inherent-conducting polymer material.

## 2. Results and discussion

### 2.1 Synthesis and characterisation of ceramic MTO

In order to improve the characteristics of *poly*-MTO with respect to its physical properties and potential usage as an inherent-conducting polymer we tried to optimise the synthetic routes of polymeric modifications of **1** to obtain a sintered ceramic material. We found that upon polymerisation and sintering of **1** at 120-140°C over a period of 10-72 hours in sealed quartz ampoules, a new material with considerably increased stability forms.[8] It does not decompose even under a pressure of up to 10 kbar, and we will therefore call it "*ceramic* MTO" **3**. Its colour and chemical composition depend on reaction time and temperature; while water-free *poly*-MTO is described as copper-coloured, with a carbon/rhenium ratio $x$ of 0.91,[6] we find varying colours (from bronze to almost black) and ratios $x$ between 0.92 (120°C, 10 h) and 0.22 (140°C, 72 h). Hence, in contrast to water-free and conventional *poly*-MTO, the stoichiometry in **3** can easily be manipulated by varying the reaction conditions. This is an interesting aspect in the material design of inherent-conducting polymers, since also the physical properties of **3** critically depend on the C/Re ratio $x$. In Fig. 1 the resistivity curves of ceramic MTO samples with different C/Re ratios $x$ of 0.68 and 0.92 are shown (the curves are normalised with respect to the one with highest methyl group concentration ($x = 0.92$)). Note that the sample containing 92 % methylated rhenium centers displays the lower residual resistivity. This is at the first glance surprising since the reduction degree of **3** formally depends on the C/Re ratio, and more of equivalents of itinerant electrons and thus a higher conductivity is expected for the more reduced species with $x = 0.68$. However, as outlined below, autoreduction of MTO during ceramic processing of **3** is accompanied by the formation of organic oxidation products which causes a non-stoichiometry of the $ReO_3$ backbone of ceramic MTO. Such oxygen deficiency apparently leads to disorder-enhanced electron localisation phenomena which again reduce the conductivity of **3**. Thus, according to our experiences a C/Re ratio around $x = 0.9$ appears to present a close to optimal stoichiometry to observe the highest possible conductivity in native ceramic MTO.[8] In the following discussion of ceramic MTO all characteristic physical and chemical properties will be related to samples of **3** displaying a C/Re ratio $x$ of 0.68 which represents a stoichiometry situated between the extreme cases of this study (C/Re ratios between $x = 0.22$ and $x = 0.92$). Hence, $(CH_3)_{0.68}ReO_{2.90}$ will be our reference compound **3** in the following sections.

**Figure 1.** The resistivity $\rho$ of the two ceramic MTO **3** samples with different C/Re ratios of $x$ = 0.68 and 0.92, respectively, is displayed by a semi-logarithmic plot vs. temperature (the solid lines are logarithmic fits between 5 K and 30 K). The resistivity follows a logarithmic temperature dependency below 30 K ($\rho = \rho_0 + A \ln((1/T))$) where the pre-factor $A$ (see ref 8 for details) increases slightly with decreasing ratios $x$.

As in **2**, the carbon content is due to the presence of methyl groups which form a covalent σ(Re-C) bond. The methyl group stoichiometry was determined (in analogy to reference 7b) quantitatively by dissolution of ceramic MTO in $H_2O_2$ and subsequent $^1$H-NMR spectroscopy of the resulting peroxo complex of MTO. The amount of methyl groups found corresponds to the carbon content determined by elemental analysis in all cases. We also determined the volatile byproducts of ceramic MTO by NMR spectroscopy and GC/MS, and found propane, ethene, water and methyl formate in addition to methane.[9] Since the only source for the oxygen contained in these byproducts is MTO, we observed a lower oxygen stoichiometry in ceramic MTO as a consequence. The oxygen/rhenium ratios determined by elemental analysis vary from 2.98 (120°C, 10 h) to 2.52 (140°C, 72 h). The formation of methyl formate is surprising at the first glance but very probably due to a Tiščenko reaction of formaldehyde. We note that formaldehyde seems to be an ubiquitous byproduct of polymerisation reactions of MTO since we could also detect it by NMR spectroscopy as major soluble constituent of the reaction mixture when MTO is polymerised in water at temperatures > 50°C. While the oxygen-bridged dimeric $Re^{VI}$ species, $(CH_3)_4Re_2O_4$, occurs as byproduct during *poly*-MTO synthesis in water,[7b,c] this compound could not be detected in the preparation of ceramic MTO.

### 2.2 ESR Spectroscopy

Since partial demethylation occurred during the synthesis of ceramic MTO, we expect the occurance of Re(VI) centres in **3**. Indeed, the presence of these paramagnetic centres in ceramic MTO could be confirmed by ESR spectroscopy (Fig. 2).[8] Two signals can be detected at 4.2 K: one (blurred by anisotropy) with six broad lines due to hyperfine coupling to rhenium ($I = 5/2$ for both $^{185}$Re and $^{187}$Re, $g = 2.0$, average $A = 0.028$ cm$^{-1}$), and a sharp singlet at $g = 2.00$ of about 1% of the total intensity. The value of 0.028 cm$^{-1}$ is at the upper limit of hyperfine coupling constants reported for monomeric Re(VI) complexes (0.0181 for $Re(CH_3)_6$ and 0.0277 for catecholate complexes).[10] In combination with detailed transport measurements one can conclude that in ceramic MTO and intercalation variants the vast majority of the localised unpaired electrons constitute Re(d$^1$) centers while only a minority of the localised spins (≈ 1%) resides at other sites of **3**. As we have demonstrated, electron localisation can be increased further by intercalation of donor molecules like tetrathiafulvalene (TTF) into **3** and by magnetic fields giving rise to a positive magneto-resistance at low temperatures (< 38 K).[8,11]

**Figure 2.** ESR spectrum of ceramic MTO **3** ($x = 0.68$) at 4.2 K showing the hyperfine structure due to coupling with nuclear spins of $^{185}$Re and $^{187}$Re ($I = 5/2$). The temperature dependence of the sharp central signal is shown in the inset.

### 2.3 NMR Spectroscopy

The structural model of ceramic MTO (see below) implies the presence of different types of methyl groups. This feature could be confirmed by $^1$H- and $^{13}$C-NMR spectroscopy. Solid state NMR spectroscopy of conducting materials may be complicated by thermally induced sample decomposition due to the heating effects resulting from the rapid sample rotation in magnetic fields. This problem was avoided by dilution of the conducting samples with KBr. This sample preparation allowed rotation frequencies up to 4 kHz. The resulting spectra revealed striking differences between **2**[12] and ceramic MTO (Fig. 3 and 4).

**Figure 3**. MAS $^1$H-NMR spectra (ppm) of ceramic MTO **3** with $x$ = 0.68 (above), spinning rate 3 kHz; *poly*-MTO **2** (below), spinning rate 4 kHz. Spinning sidebands (residual chemical shift anisotropy) can be seen in both spectra.

**Figure 4**. CP/MAS $^{13}$C-NMR spectra (ppm) of ceramic MTO **3** with $x$ = 0.68 (above), spinning rate 3 kHz; *poly*-MTO **2** (below), spinning rate 4 kHz.

In the $^1$H-NMR spectrum of ceramic MTO **3** (Fig. 3, above) the central signal is composed of four not completely resolved lines at -2.0, -2.5, -3.0 and -3.4 ppm of similar intensity. There is a broad hump at about 0 ppm of low intensity. In contrast, **2** (Fig. 3, below) shows a main signal centered at -1 ppm, and a second one at 8 ppm, in a ratio of about 5:1. Considering chemical shift arguments, we attribute the upfield signal to methyl groups and the downfield signal to water molecules separating the layers. This leads to an average methyl/water ratio of approximately 3:1.[13] Compared with solid monomeric MTO, the mobility of the methyl groups in *poly*-MTO and ceramic MTO appears to be considerably reduced, and thus no broad sidebands due to $^1$H-$^1$H dipolar coupling are observed, even at low spinning rates (500 Hz). The structure of the $^{13}$C-NMR signals is also strikingly different for both samples. In ceramic MTO **3** (Fig. 4, above) one observes two partially overlapping signals of similar intensity centered at 47.3 and 27.5 ppm. In addition, there is a small sharp peak of low intensity at 3.5 ppm which can not be related to the presence of (unreacted) residual monomeric MTO ($\delta$ = 36 ppm). In contrast, the $^{13}$C-NMR spectrum of **2** (Fig. 4, below) presents two very distinct broad peaks at 40 and -44 ppm (ratio 4:1). While the chemical shift of the downfield signal corresponds well to ceramic MTO, the large upfield shift of the minor signal is striking. A tentative explanation would be that *poly*-MTO contains domains with a higher *local* concentration of paramagnetic centres (upfield signal) while ceramic MTO is more homogeneous in this respect.

### 2.4 IR Spectroscopy

According to the structural model of ceramic MTO, rhenium-oxygen double bonds should be present. Clear evidence for the presence of intact Re=O and Re-C bonds can be derived from IR spectroscopy studies (Fig. 5). For **1** in an argon matrix at 14 K, three salient signals were observed (970.3 (vs); 1000.7 (w) and 565 cm$^{-1}$ (w)) and assigned to the asymmetric and symmetric ReO$_3$ stretching and Re-C stretching modes, respectively.[14] These modes shift to slightly different values in the case of solid **1** ($\nu_{as}$(ReO$_3$) = 956/963 (vs); $\nu_s$(ReO$_3$) = 998 (w) and $\nu$(Re-C) = 569 (w) cm$^{-1}$; Fig. 5). *Poly*-MTO **2** and ceramic MTO display related modes which are, however, shifted to lower wave numbers; **2** (CsI): 913 (vs), 955 (w) and 549 (vw) cm$^{-1}$; ceramic MTO **3** (KBr): 910 (vs), 940 (m) and 501 (vw) cm$^{-1}$, respectively.[8] The shift indicates a weakening of the Re=O and Re-C bonds during the formation of ceramic MTO, in accord with the expectations inferred from the structural model (see below).

**Figure 5.** Comparison of the Re=O and Re-C stretching modes revealed by the IR spectra of **1** and **3** ($x$ = 0.68).

### 2.5 In-situ Diffraction Studies

The temperature and time dependence of the polymerisation and sinter processes during formation of ceramic MTO was studied by *in-situ* diffraction and compared with the aqueous polymerisation of MTO leading to **2** (see Experimental Section for details).

For the formation of **2** oversaturated aqueous dispersions of MTO were studied in a capillary, and the temperature was raised from 50°C to 90°C employing a heating rate of 4°C/min. At the beginning of the temperature interval the diffraction pattern showed basically Bragg reflections of still undissolved MTO single crystals. At about 80°C a colour change of the MTO solution from colourless to blue was observed which marks the onset of the polymerisation reaction. The polymerisation process is apparently faster than the time resolution of our X-ray experiment (one diffraction pattern/180 s) since the colour change of our solution was immediately accompanied by the appearance of weak but characteristic powder rings of **2**.[7c] Longer reaction times solely lead to a sharper and more intense powder diffraction pattern but do not reveal any other crystalline byproducts of **2**. After a reaction period of 60 minutes no significant change of the diffraction pattern was noticeable and the final diffraction pattern depicted in Fig. 6 shows solely the Bragg intensities of **2** (Fig. 6).[15]

Under non-aqueous reaction conditions formation of ceramic MTO was studied *in situ* in a similar way. Here, the temperature was raised up to 120°C using a rate of 1°C/min. At 117°C MTO is completely molten, as indicated by the vanishing of all Bragg peaks of crystalline MTO. The first sign for the start of the polymerisation process was indicated by subtle but characteristic powder diffraction rings of **3** which appeared after a tempering period of 30 minutes at 120°C. Again, the formation of powder rings is accompanied by a colour change of the melt from colourless to blue. After a time span of three hours at 120°C the intensity and sharpness of the powder diffraction pattern of **3** increased only slowly (see the final pattern in Fig. 6). Hence, in contrast to **2** longer reaction times do not significantly improve the crystallinity of ceramic MTO.

**Figure 6.** Final diffraction pattern of **2** and ceramic MTO **3** formed under *in-situ* reaction conditions. **2** has been indexed by an orthorhombic unit cell ($a = b = 3.728(1)$, $c = 16.516(5)$ Å; *P4mm*)[7c] while **3** is characterised by a two-dimensional lattice of *p4mm* symmetry ($a = 3.66(1)$ Å).

### 2.6 Structural model of ceramic MTO and *poly*-MTO

The diffraction pattern obtained for **2** and ceramic MTO **3** reveal a close structural relationship of both compounds with respect to the formation of extended $\{ReO_2\}_\infty$ layers (Fig. 6), as revealed by comparable diffraction angles of the *hk*0 Bragg peaks. Previous X-ray studies[7c] of **2** led to a structural model displaying double layers of corner-sharing $[CH_3ReO_5]_\infty$ octahedra interconnected by intercalated water molecules. However, no such 3D ordering can be expected for ceramic MTO since only the *hk*0 reflection series are evident in the diffraction pattern (Fig. 6). Apparently, crystallographic ordering in ceramic MTO occurs exclusively in two dimensions. This finding is further supported by the pronounced asymmetry of the reflection profiles of ceramic MTO (Fig. 6) which show a slow decay of Bragg intensity in the direction of increasing diffraction angles (*cf.* 100 and 110 reflection in Fig. 6). This peak shape asymmetry is a well-known indicator of layered compounds displaying a turbostratic or 00*l* defect stacking.[7c] We note that the same peak shape asymmetry is also evident for **2** but less pronounced due to the presence of intercalated water molecules which serve as linkers of adjacent oxide layers via hydrogen bonding [Re=O···H-O-H···O=Re].[7c] The reciprocal space reconstruction for the resulting 2D diffraction pattern of **3** therefore consists of diffuse 00*l* rods spread parallel to the *c* axis (stacking direction). Hence, the formation of diffuse 00*l* rods is responsible for the smooth decay of the Bragg intensity of each *hk*0 reflection with increasing diffraction angles.

The purely 2D ordering of the ceramic MTO is further supported by the physical properties displayed by this compound:[8] The resistivity data of pure ceramic MTO exhibit a crossover from metallic ($d\rho/dT > 0$) to insulating ($d\rho/dT < 0$) behaviour at a characteristic temperature around $T_{min} \approx 38$ K. Above $T_{min}$ the resistivity $\rho(T)$ is remarkably well described by a two-dimensional electron system. Below $T_{min}$ an unusual resistivity behaviour, similar to that found in doped cuprates (e.g. $La_{1.85}Sr_{0.15}Zn_yCu_{1-y}O_4$)[16], is observed: The resistivity initially increases approximately as $\rho \sim \ln(1/T)$ before it changes into a $\sqrt{T}$ dependence below 2 K (Fig. 1). As an explanation we suggest a crossover from purely two-dimensional charge-carrier diffusion within the $\{ReO_2\}_\infty$ planes at high temperatures to three-dimensional diffusion at low temperatures in a disorder-enhanced electron-electron interaction scenario[8] (Altshuler-Aronov correction[17]).

Taking the indexing results of the X-ray diffraction study into account, it is reasonable to assume – like for *poly*-MTO - that the $\{ReO_2\}_\infty$ units span the 2D unit cell of *p4mm* symmetry characterised by the lattice parameter $a = 3.66(1)$ Å. The lattice parameter $a$ is related to but significantly smaller than those reported for the cubic inorganic oxide $ReO_3$ ($a = 3.748(1)$ Å),[7c,18] the mixed oxides $Re_xW_{1-x}O_3$ ($a = 3.7516(2)$; $x = 0.25$)[4] and cubic $WO_3$ (3.7719(4) Å).[19] We note that the unit cell parameters in $Re_xW_{1-x}O_3$ can be used as sensitive measure of the Re/W ratio: isotypic replacement of $Re^{VI}$ by $W^{VI}$ ions leads to a unit cell expansion in accordance with the different ionic radii of 0.55 and 0.60 Å (coordination number 6), respectively.[20] In the same line we might argue that the smaller unit cell parameter of ceramic MTO relative to $ReO_3$ are due the presence of (smaller) $Re^{VII}$ vs. $Re^{VI}$ centers in ceramic MTO.

Due to the close structural relationship to the inorganic parent compound $ReO_3$ (spacegroup *Pm-3m*) an idealised structural model for ceramic MTO can be derived by adopting its perovskite structure in two dimensions in form of $\{ReO_2\}_\infty$ layers. Complementing the coordination environment of the Re atom by one methyl and one oxo group leads to a layered network of $[CH_3ReO_5]_\infty$ octahedra displaying an averaged Re–Re separation of about 3.66 Å. In order to account for the missing methyl groups in *poly*-MTO $(CH_3)_xReO_3$, this model can be further developed by an idealised superstructural model consisting of three MTO and one demethylated $ReO_3$ unit: $[CH_3ReO_3]_3[ReO_3]$ **3a** (Fig. 7a). This model leads to a ratio C/Re of 0.75 which is well within the range of methyl group content obtained under various experimental conditions ($x = 0.22$-$0.92$). Such a model would rule out any significant interlayer interactions and is supported by specific heat capacity studies in a wide temperature range over more than three temperature decades (0.08 K – 300 K).[8] According to these studies all layers of ceramic MTO are nearly ideally decoupled and solely connected by subtle van der Waals interactions. This fact is reflected particularly by the large 2D phonon contribution in specific heat measurements.[8,11a] Furthermore SEM micrographs of ceramic MTO show the rather different morphology of *poly*-MTO and its ceramic variant. The latter displays the typical features of a partially amorphous sintered material while its polymeric congener clearly shows the characteristic morphology of a layered oxide (Fig. 8).

**Figure 7.** (a) Fragment of ceramic MTO based on an optimised $\{[ReO_3][CH_3ReO_3]_3\}_n$ superstructural model **3a** employing DFT calculations and 2D periodic boundary conditions. Selected bond distances [Å] and angles [deg]: Re1-Re2 3.77, Re2-Re3 3.81, Re3-Re4 3.81, Re4-Re1 3.78, Re1=O1 1.74, Re1-O5 1.81, Re1-O6 1.81, Re1-O7 1.96, Re1-O11 1.96, Re2-C2 2.19, Re2=O2 1.77, Re2-O7 1.86, Re2-O8 2.02, Re2-O9 2.08, Re2-O10 1.80, ∠C2-Re2-

O2 149.2. (b) Isosurface map of the negative Laplacian of the electron density $-\nabla^2\rho(r)$, around Re2 at the contour value of 263 eÅ$^{-5}$.

Optimisation of our idealised model of **3a** of ceramic MTO employing periodic boundary conditions and the PBEPBE/CRENBL/3-21G(d) approximation highlights another reason for the partial amorphous character displayed by ceramic MTO. Geometry optimisations lead to a highly distorted octahedral geometry as signalled by average ∠C-Re=O angles of 149.7° and a pronounced bond length asymmetry in the {ReO$_2$}$_\infty$ planes: two short Re-O distances (1.82 Å) and two long Re-O distances (2.05 Å) at each methylated rhenium atom; averaged values, respectively (Fig. 7). This bond length asymmetry is somewhat reduced in the case of the demethylated metal centers but the two short (1.81 Å) / two long (1.96 Å) patterning of the equatorial Re-O distances is still preserved at the formal Re($d^1$) centers.[21] We suggest that the origin of these distortions can be correlated with the presence of pronounced ligand-induced charge concentrations (LICC)[22] in the valence shell of the methylated rhenium centers (Fig. 7b). Hence, a linear arrangement of the C-Re=O structural moiety is energetically destabilised due to Pauli repulsion between the electron pair domains of the strong axial Re-C/Re-O bonds and their opposing ligand-induced charge concentrations denoted LICC(C) and LICC(O) in Fig. 7b. As another consequence of the energetically destabilised C-Re=O units the Re-C and to a lesser extent the axial Re-O bonds in ceramic MTO are elongated and thus activated compared with MTO: Re-C = 2.17/2.063(2) Å and Re-O = 1.77/1.702(1) Å in ceramic MTO and solid MTO,[7c,23,24] respectively (average values for ceramic MTO).

**Figure 8**. SEM micrographs of **2** (left) and ceramic MTO **3** (right).

These characteristic structural features are hence in fine agreement with our spectroscopic studies. Fig. 5 shows the dominant vibrational double mode at 910 and 940 cm$^{-1}$ of **3** in the characteristic region of Re=O stretching modes and a weak feature at 501 cm$^{-1}$ due to the ν(Re-C) streching frequency.[8] These modes are clearly weakened in comparison with the corresponding modes in solid (monomeric) MTO (Fig. 5). Hence, the shift of the vibrational excitations to lower energies is in fine agreement with the above predicted Re=O and Re-C bond elongation of about 0.07 and 0.11 Å, respectively, relative to solid MTO. Furthermore, solid state $^{13}$C and $^1$H MAS NMR studies of ceramic MTO confirm the presence of methyl groups in different chemical environments in conformity with the superstructural DFT model **3a** (see Fig. 3 and 4).

As outlined above the significantly activated Re-C bond in ceramic MTO is thus a direct consequence of the hexa-coordination of rhenium in the extended structure of ceramic MTO. On one hand the structural distortions reflect an energetic destabilisation of the Re-C and Re=O bonds relative to MTO; on the other hand they facilitate the elimination of methyl groups and are therefore the true driving force allowing the formation of an extended ceramic solid. Accordingly, the energetic destabilisation of the localised chemical bonds in ceramic MTO is more than compensated by the win in kinetic energy due to the formation of a conduction band hosting delocalised itinerant electrons. In the next chapter we will outline that the dramatic changes in the electronic structure of MTO during the formation of a ceramic material also leads to new and unexpected chemical reactivity.

**Perspectives**

We have shown in the previous sections how to synthesise ceramic MTO by a sintering process from monomeric MTO and determined its chemical composition and physical properties. We also demonstrated that it is comparatively easy to modify the stoichiometry of ceramic MTO and to manipulate its physical properties by intercalation of donor or acceptor molecules during the sintering process;[8] such controlled intercalation is considerably more difficult to achieve during synthesis of **2** in aqueous solution. DFT calculations indicate that ceramic MTO has structural features different from monomeric MTO, *e.g.* activated Re-C and weaker Re=O bonds. Hence, we can expect different chemistry for the ceramic modification of MTO. For example, the polymerisation process of MTO **1** appears to be an efficient starter for the polymerisation of tetrathiafulvalene (TTF) which does not form polymers under typical radical or ionic polymerisation conditions nor with MTO in solution.[8] Indeed, attempts to intercalate TTF during the polymerisation and subsequent sintering process of ceramic MTO at higher TTF/Re ratios (> 1) did not lead to isolated intercalated TTF guest molecules but rather to a solid solution of polymeric MTO in an organic polymeric TTF matrix. Apparently, the polymerisation of MTO initiates the copolymerisation process with TTF.

As another example, we note that reactions of ceramic MTO with aliphatic amines and ammonia do not lead to simple base adducts as can be obtained from Lewis-acidic monomeric MTO in diluted solutions,[1] but to intercalation products similar to those known from layered metal dichalcogenides, e.g. $TaS_2$.[25]

Redox reactions are likely to accompany such intercalations with amines. Furthermore, with amines of the general structure $R^1R^2CH-NH_2$, ceramic MTO displays a reactivity which is not shown by the monomer itself. Thus, cyclohexylamine and 1-phenylethylamine react with ceramic MTO under the formation of imines $R^1R^2C=N-CHR^1R^2$, i.e. with oxidation and alkyl transfer. Activated Re=O bonds in ceramic MTO are likely to be involved in this reaction. A possible mechanism is shown in Fig. 9.

**Fig. 9**. Possible two-step mechanism for the formation of $R^1R^2C=N-CHR^1R^2$ from ceramic MTO and amines $R^1R^2CH-NH_2$.

Since these imines are formed by oxidation and alkyl transfer reaction, we obtain a modified ceramic MTO which contains intercalated amine and ammonia; typical compositions are $(CH_3)_{0.8}ReO_{3-y}(C_6H_{13}N)_{0.13}(NH_3)_{0.13}$ for cyclohexylamine and $(CH_3)_{0.8}ReO_{3-y}(C_8H_{11}N)_{0.06}(NH_3)_{0.17}$ for 1-phenylethylamine, where $y$ accounts for the understoichiometry of oxygen in the starting ceramic MTO **3** in addition to the oxygen consumed by the oxidation of the amine to form the imine (typically $y < 0.2$). Monomeric MTO forms a normal adduct with 1-phenylethylamine in toluene solution but behaves differently when treated with the neat amine at room temperature. A rapid darkening with precipitation of a black material sets in, and the corresponding imine can be detected in the reaction mixture. This can readily be understood in terms of redox chemistry: First, MTO reacts with the amine under liberation of ammonia and acetophenone and the formation of a reduced rhenium centre. This acts as starting center for the polymerisation reaction. The polymer then induces the formation of the imine. When the solid product is isolated after several days of reaction, it shows similar properties and composition as the one obtained directly from ceramic MTO and 1-phenylethylamine. This unprecedented oxidation behaviour of **3** warrants further exploitation.

**Experimental**

**Resistance measurements**: These studies were performed using a four-point low-frequency ac-method in the temperature range between 30 mK and 300 K. For the determination of the resistivity of ceramic MTO one has to take into account a rough estimate of the non-uniform sample shape of **3** causing a relative error bar of 30% for the residual resistivity.

**In-situ diffraction study:** A glass capillary (1.0 mm diameter) was loaded with approximately 6 mg MTO **1** and 3 μl water to study the formation of **2**. For the *in-situ* formation study of **3** we used a glass capillary (0.5 mm diameter) filled with about 1 mg MTO, respectively. In both cases the capillaries were mounted on an Eulerian cradle (Huber) and exposed to monochromated Mo-$K_\alpha$ radiation of a FR591 rotating anode (Bruker AXS). Diffraction pattern were recorded using a MAR345 area imaging detector which covered the resolution range from $1.03 < d < 9.69$ Å$^{-1}$. Preferred orientation effects of samples were reduced by sample rotation. The temperature conditions were controlled by an $N_2$ open-flow heating device (*cf.* Fig. 6).

**Computational study of 3:** The geometries of several two-dimensional models of ceramic MTO were fully optimised using density functional theory (DFT) methods with periodic boundary conditions as implemented in the GAUSSIAN03 program package.[26] The PBEPBE functional in combination with the CRENBL basis-set and an averaged relativistic effective core potential for Re and a standard 3-21G* basis-set for C, H and O was used throughout.[27] Auxiliary density fitting functions were generated employing the standard algorithm implemented in GAUSSIAN03.

**Analysis and spectroscopy:** NMR spectra were obtained with a Varian MercuryPlus 400 spectrometer (proton frequency 400.42 MHz). Liquid state spectra were performed in CDCl$_3$ unless stated otherwise. Solid state NMR spectra were obtained from samples diluted with KBr (60-80 mg in 200 mg KBr) by conventional MAS and CP-MAS (for $^{13}$C) techniques. IR spectra were recorded on a Thermo Nexus FT-IR spectrometer with an ATR unit on neat samples, while transmission spectra were obtained in CsI or KBr pellets at room temperature on a Bruker IFS113v with a Si-bolometer as detector for the range of 600 to 100 cm$^{-1}$ and a MCT detector for the range of 4000 to 500 cm$^{-1}$. The content of Re was determined with a Varian ICP-OES Vista MPX Simultanous. C, H, N and O analyses were done with an ELEMENTAR elemental analyser vario EL III. ESR spectra were recorded at X-band frequencies (9.4 GHz) with a Bruker ELEXSYS E500-CW spectrometer using a continuous helium gas-flow cryostat (Oxford Intruments) for temperatures between 4.2 and 300 K.

**Methyltrioxorhenium 1**: MTO was prepared as described.[28] All other reagents were commercially available and used as received. Solvents were dried by conventional techniques and stored over molecular sieves (3 Å). The content of methyl groups in samples of ceramic MTO was determined as described[7b] by $^1$H-NMR spectroscopy (dissolution in a $H_2O_2/D_2O$ mixture and comparison of the integrals over the peroxo complex [Re(CH$_3$)(O$_2$)$_2$(O)] and methanol with a known amount of acetonitrile).

**Ceramic MTO 3**: The appropriate amount of MTO (50-1000 mg) was heated in a sealed quartz ampoule under vacuum to the appropriate temperature (120-140°C) for the desired time (10-72 h). The reaction is complete under these conditions since no monomeric MTO can be extracted from ceramic MTO by organic solvents (hexane, toluene, or diethyl ether). The yield is > 90%. The colour of the materials varies from golden (120°C) over copper-coloured to almost black (140°C). Typical results for different reaction conditions are listed below. The formulae were calculated from analytical data (C, H, O: elemental analysis, Re: ICP determination). Physical and spectroscopic measurements were performed with the sample of run 2, and chemical reactions with the sample from run 1.

| run | temp. (°C) | time (h) | composition |
|-----|------------|----------|-------------|
| 1 | 120 | 10 | $(CH_3)_{0.92}ReO_{2.98}$ |
| 2 | 120 | 48 | $(CH_3)_{0.68}ReO_{2.90}$ |
| 3 | 130 | 72 | $(CH_3)_{0.44}ReO_{2.70}$ |
| 4 | 140 | 72 | $(CH_3)_{0.22}ReO_{2.52}$ |

Volatile byproducts of the formation of ceramic MTO were trapped by performing the reaction in a flask with a rubber septum under argon and transferring the gaseous phase by a steel canule into a screw cap NMR tube containing $CDCl_3$ precooled with liquid nitrogen. The identity of the products (methane, propane, ethene, methyl formate, water) was ascertained by two-dimensional NMR techniques (COSY, HSQC, HMBC).

**Reaction of ceramic MTO with 1-phenylethylamine and cyclohexylamine**: Ceramic MTO **3** ($x = 0.92$; 100 mg) and the amine (1 mmol) are stirred under nitrogen at room temperature for 5 d. The mixture is then extracted with diethyl ether (3 times 1 ml). The ether solution contains the corresponding imine (N-cyclohexylidene-cyclohexylamine or N-(1-phenylethylidene)-1-phenylethylamine, respectively), identified by NMR spectroscopy, in addition to excess starting amine (typical imine/amine ratio 0.07). The solid residue was analyzed by elemental analysis and IR spectroscopy. Product from cyclohexylamine: N: 1.22 %; C: 7.16 %; H: 1.94 %; Re: 69.6 %;  IR (CsI): 3511 (m, ν(NH)), 3019 (w), 2940 (m, ν(CH)), 2860 (w), 1612 (m), 1491 (w), 1459 (w), 1410 (m); 1118 (w), 1064 (w), 1031 (w), 1009 (w); 970 (w), 912 (vs, ν(ReO)), 337 (w), 320 (m) cm$^{-1}$. Product from 1-phenylethylamine: N: 1.22 %; C: 5.89 %; H: 1.11 %; Re: 72.1 %; IR (CsI): 3456 (m, ν(NH)), 3222 (w), 3030 (w),  1613 (m), 1410 (s), 1031 (m), 913 (vs, ν(ReO)), 768 (w), 689 (w), 633 (w), 319 (w) cm$^{-1}$.

**Reaction of 1-phenylethylamine with neat MTO 1**: MTO **1** (108 mg, 0.42 mmol) and 1-phenylethylamine (145 mg, 1.19 mol) are stirred under nitrogen at room temperature. The mixture turns dark after 10 minutes and forms a semisolid after 30 minutes. The reaction is left stirring for 2 d. Workup is done as described for the reaction with ceramic MTO. The ether solution shows an imine/amine ratio of 0.06. The black solid residue (105 mg) shows N, 4.84%; C, 13.68%; H, 3.95%, Re, 62.3%.

**Acknowledgement**

This work was supported by the Deutsche Forschungsgemeinschaft (DFG) through the SFB 484 and by Nanocat, an International Graduate Program within the Elitenetzwerk Bayern.

**Figure 1**
**Click here to download high resolution image**

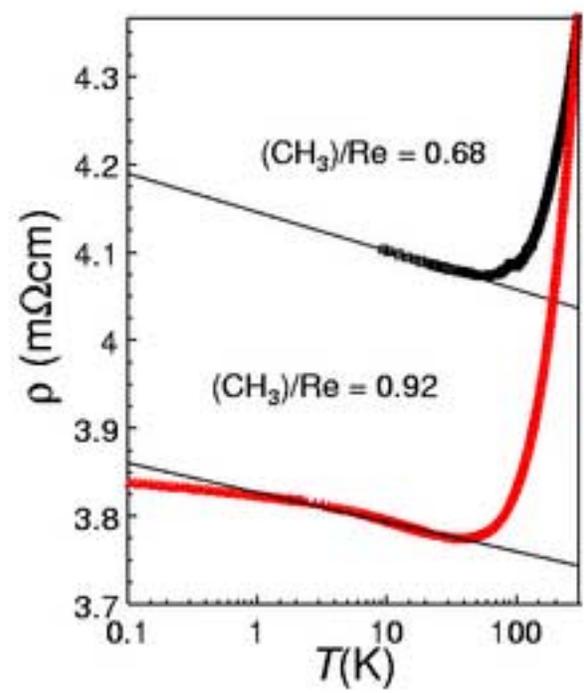

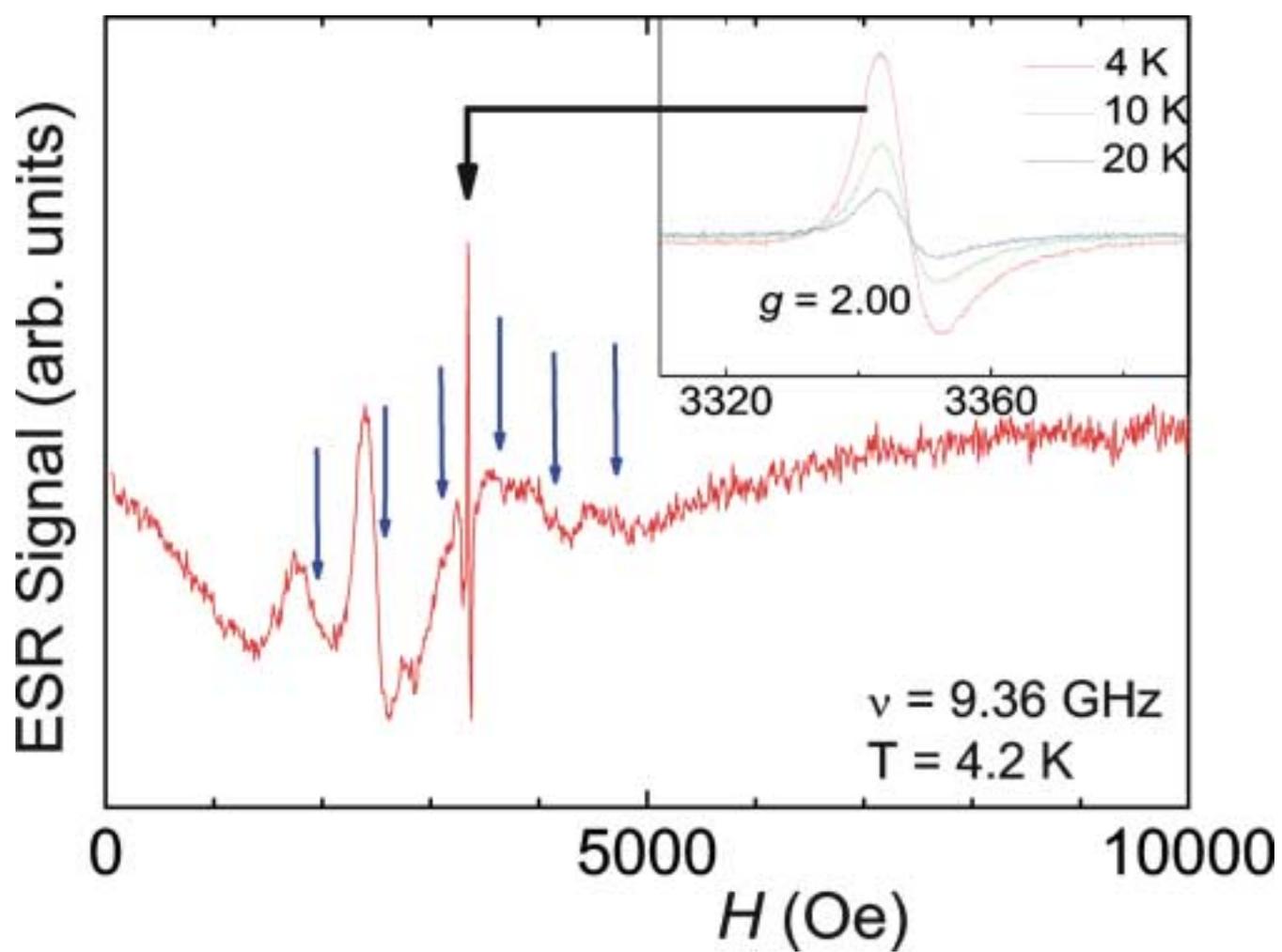

**Figure 3**
**Click here to download high resolution image**

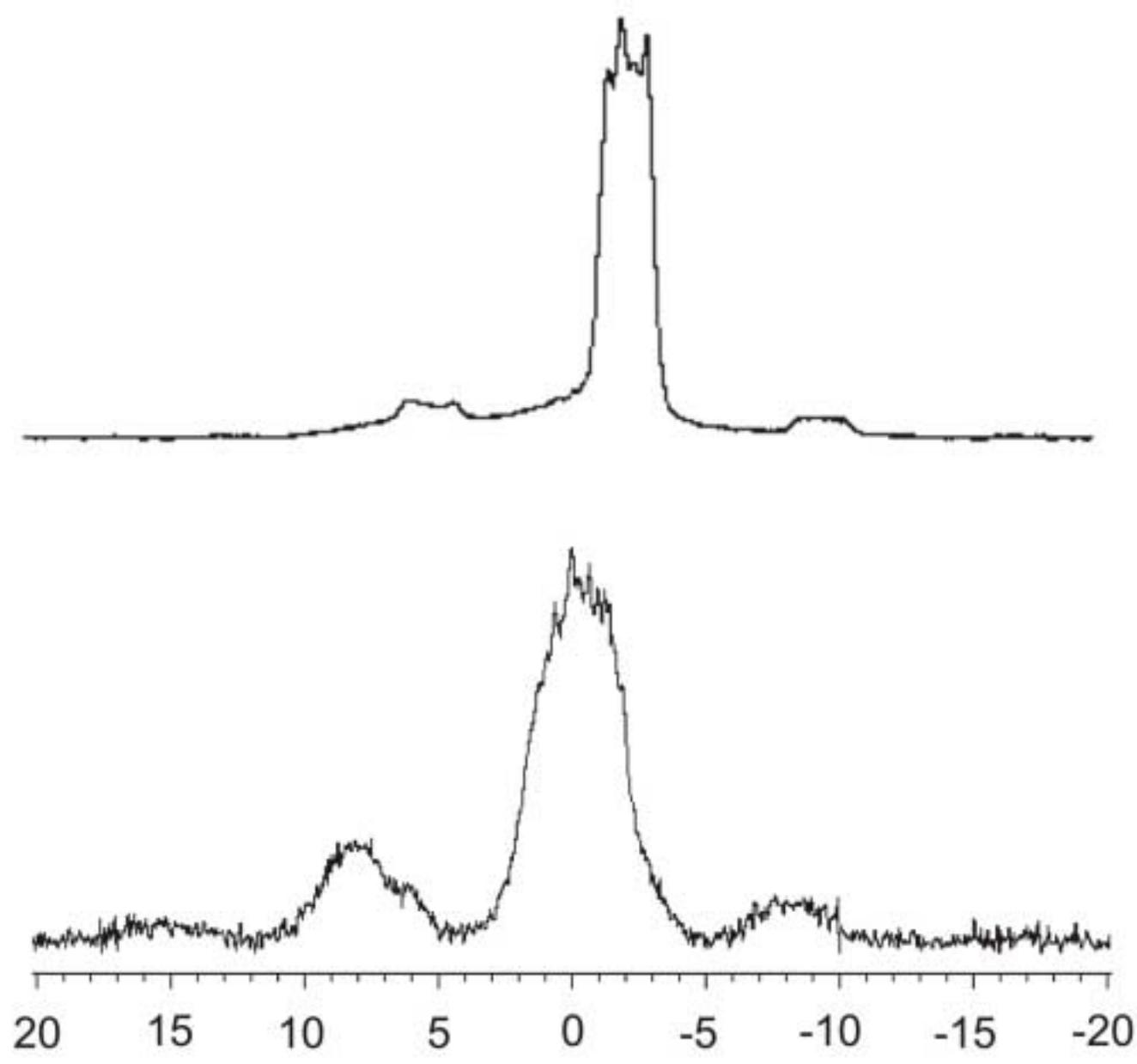

**Figure 4**
**Click here to download high resolution image**

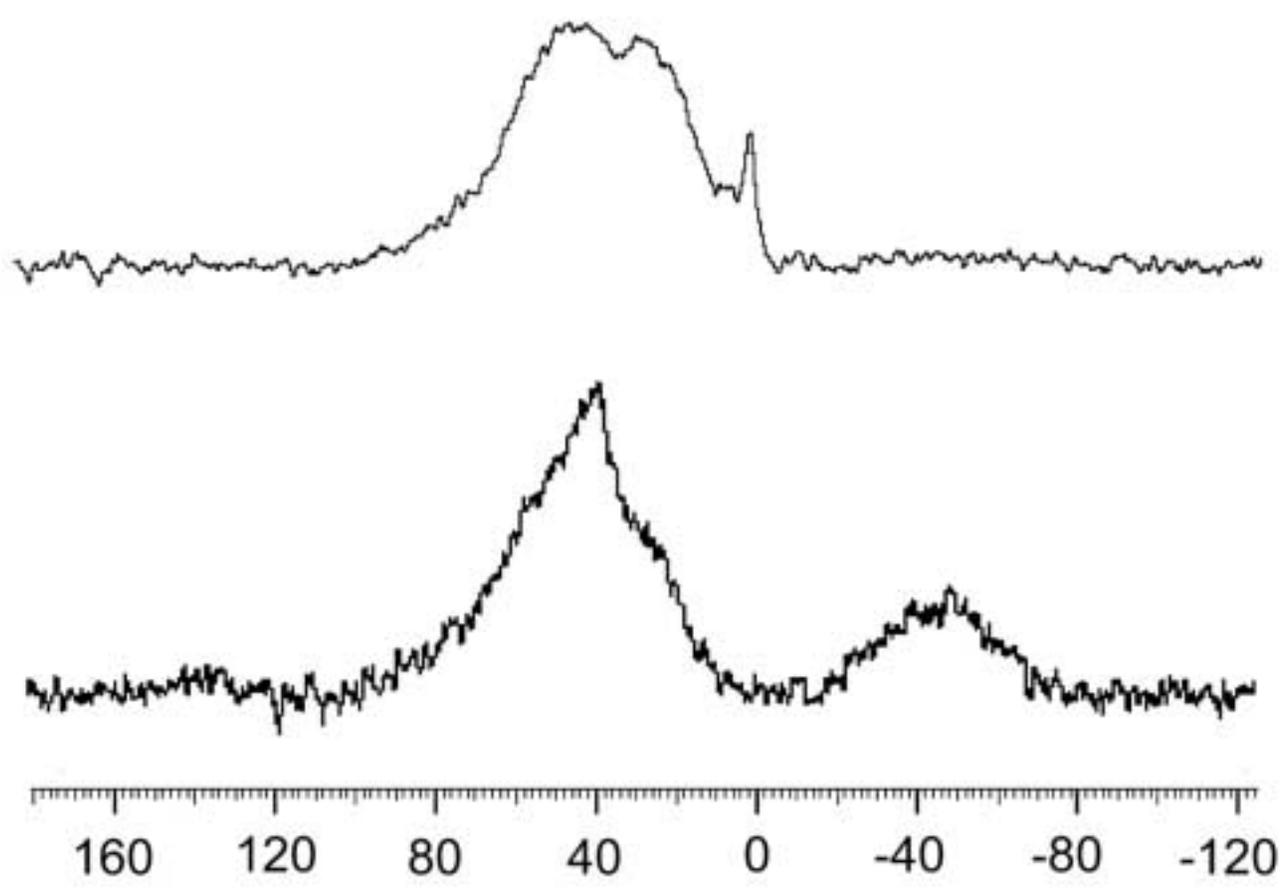

**Figure 5**
Click here to download high resolution image

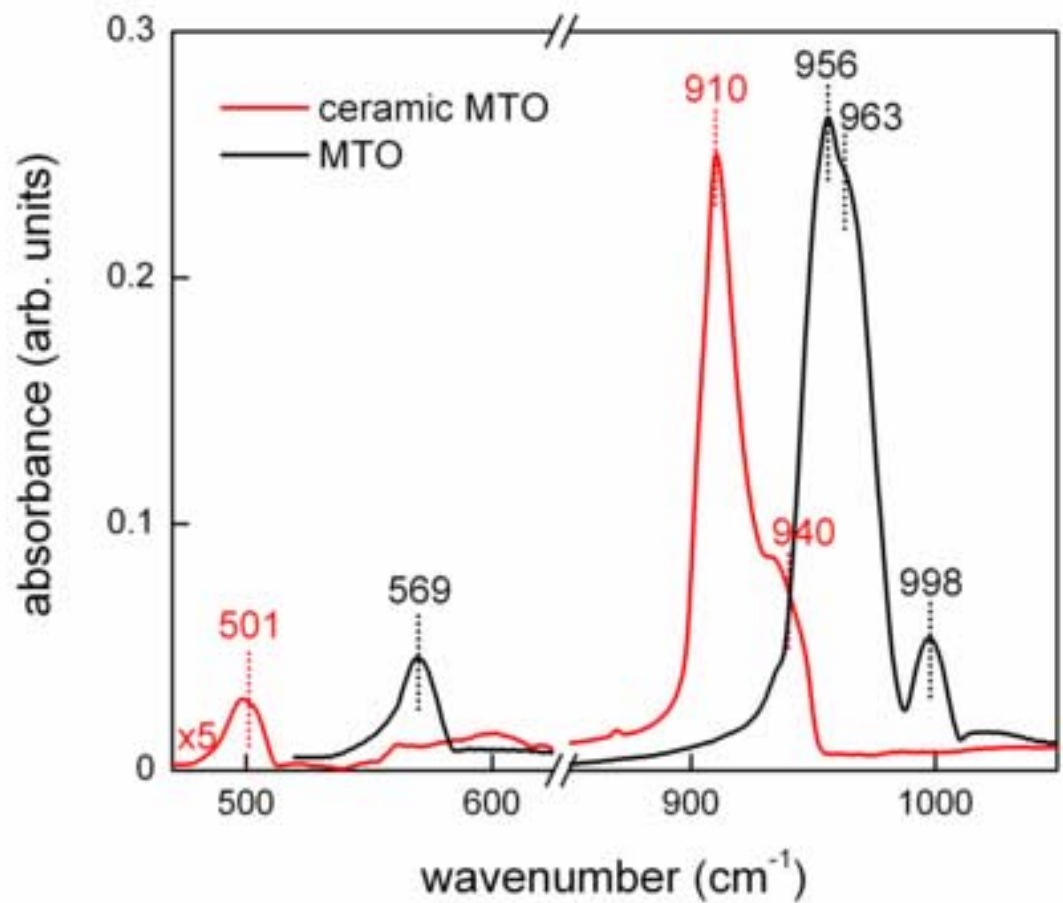

**Figure 6**
**Click here to download high resolution image**

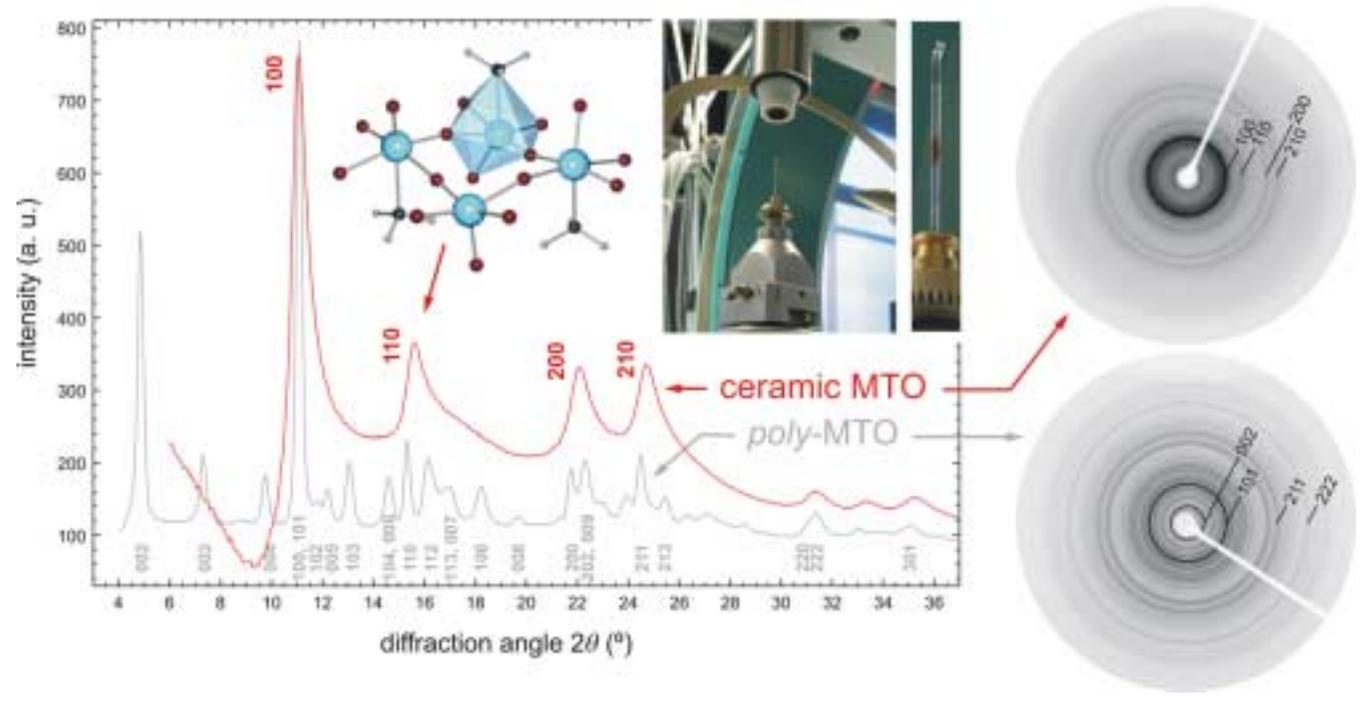

**Figure 7**
**Click here to download high resolution image**

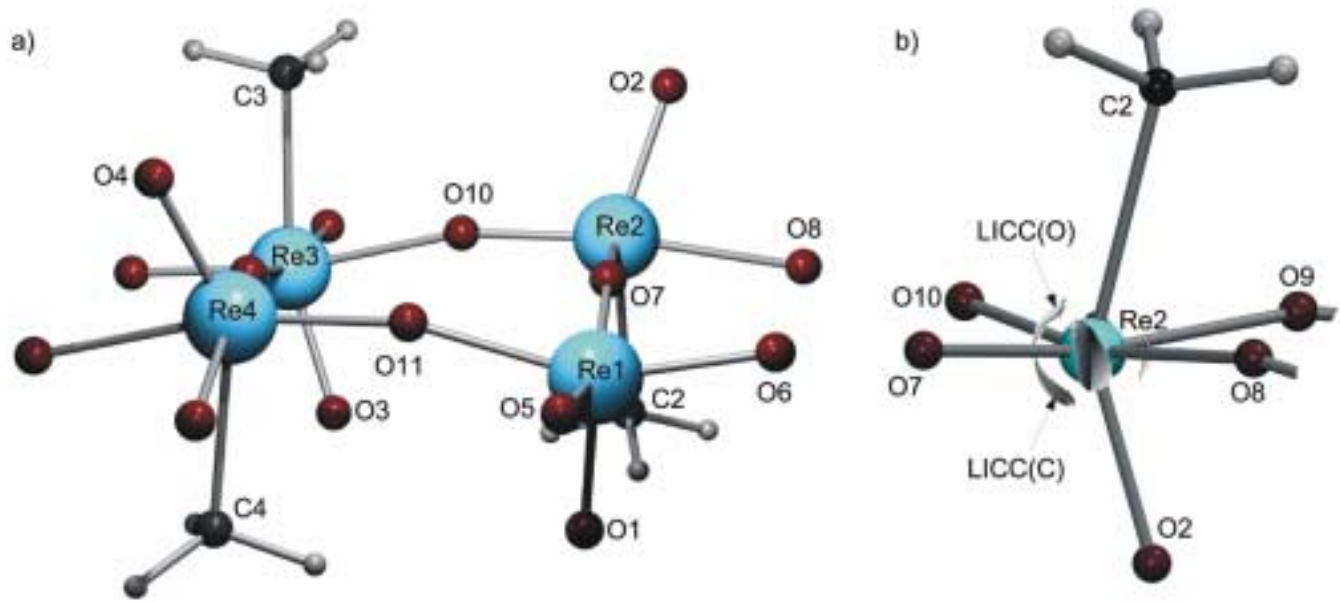

**Figure 8**
**Click here to download high resolution image**

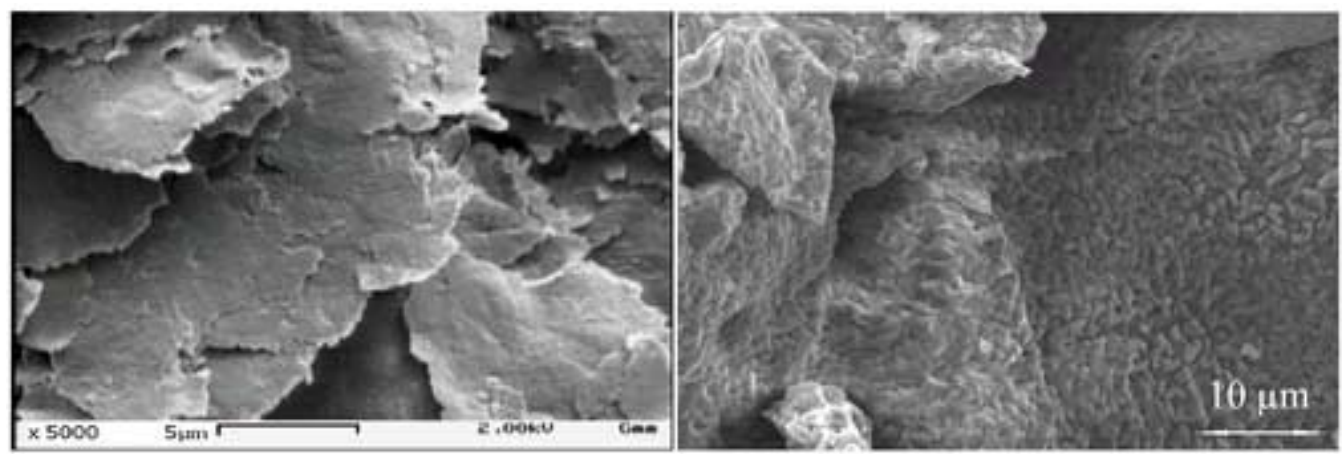

**Figure 9**
**Click here to download high resolution image**

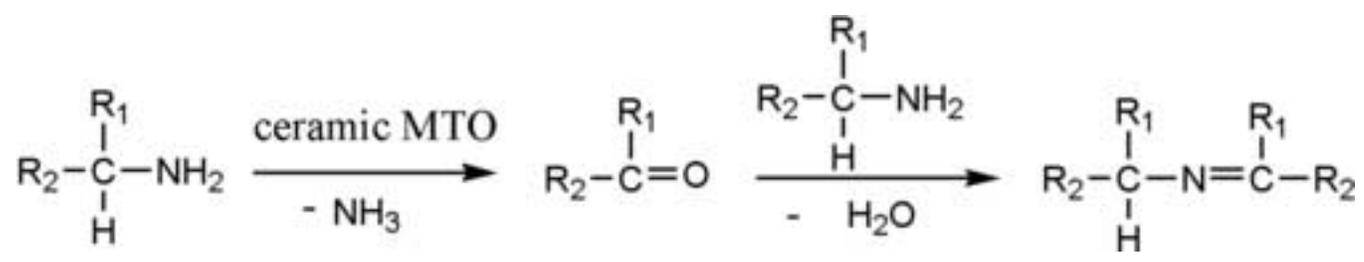

**Graphical Abstract (pictogram)**

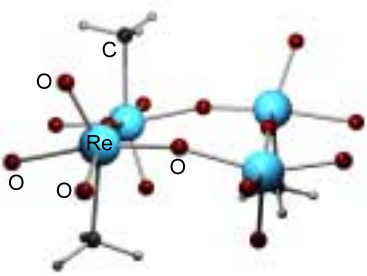

**Graphical abstract (synopsis)**

**Graphical Abstract**

**Ceramic Methyltrioxorhenium**

Rudolf Herrmann, Klaus Tröster, Georg Eickerling, Christian Helbig, Christoph Hauf, Robert Miller, Franz Mayr, Hans–Albrecht Krug von Nidda, Ernst-Wilhelm Scheidt, Wolfgang Scherer*

Chemische Physik und Materialwissenschaften, Universität Augsburg, Universitätsstr. 1, D-86159 Augsburg, Germany

The metal oxide polymeric methyltrioxorhenium $[(CH_3)_xReO_3]_\infty$ (*poly*-MTO) is an unique representative of an inherent conducting organometallic polymer. To improve its physical and mechanical characteristics to we propose a solvent-free synthesis of *poly*-MTO leading to a sintered ceramic material, denoted ceramic MTO.